\documentclass[aps,pre,reprint,amsmath,onecolumn,amssymb,groupaddress,superscriptaddress]{revtex4-2}

\usepackage[utf8]{inputenc}
\usepackage{amsmath,amsfonts,amssymb,mathtools,mathrsfs}

\usepackage{bm}
\usepackage{color}
\usepackage{graphicx}
\usepackage{soul}
\usepackage{CJK}
\usepackage{xcolor}
\usepackage{appendix}
\usepackage{tikz}
\usetikzlibrary{positioning, arrows.meta, backgrounds, calc, fit}
\usepackage{cancel}

\usepackage{tikz}

\newcommand{\CW}{{\rm CW }}
\newcommand{\CCW}{{\rm CCW }}

\newcommand{\CWX}{ {\rm CW}_x}
\newcommand{\OCWX}{{\overline {\rm CW}_x}}
\newcommand{\CCWX}{ {\rm CCW}_x}
\newcommand{\OCCWX}{{\overline {\rm CCW}_x}}

\begin{document}

\begin{CJK*}{UTF8}{gbsn}

\title{Non-equilibrium symmetry of cyclic first-passage times}
\author{Daniel M. Busiello}
\email{danielmaria.busiello@unipd.it}
\affiliation{Department of Physics and Astronomy ``G. Galilei'', University of Padova, 35131 Padova, Italy}
\affiliation{Max Planck Institute for the Physics of Complex Systems, 01187 Dresden, Germany}
\affiliation{Theoretical Sciences Visiting Program (TSVP), Okinawa Institute of Science and Technology Graduate University, Onna, Okinawa 904-0495, Japan}
\thanks{The authors are listed in alphabetical order.}
\author{Shiling Liang}
\email{shiling@pks.mpg.de}
\affiliation{Max Planck Institute for the Physics of Complex Systems, 01187 Dresden, Germany}
\affiliation{Biological Complexity Unit, Okinawa Institute of Science and Technology Graduate University, Onna, Okinawa 904-0495, Japan}
\affiliation{Center for Systems Biology Dresden, 01307 Dresden, Germany}
\affiliation{Max Planck Institute of Molecular Cell Biology and Genetics, 01307 Dresden, Germany}
\thanks{The authors are listed in alphabetical order.}
\author{Simone Pigolotti}
\email{simone.pigolotti@oist.jp}
\affiliation{Biological Complexity Unit, Okinawa Institute of Science and Technology Graduate University, Onna, Okinawa 904-0495, Japan}
\thanks{The authors are listed in alphabetical order.}

\begin{abstract}
We study the sum of first passage times along an arbitrary cycle made up of $N>2$ states of a small physical system. We show that, if the system is at thermodynamic equilibrium, this sum follows the same probability distribution regardless of whether the cycle is explored clockwise or counterclockwise. Out of equilibrium, the distributions of clockwise and counterclockwise cyclic first passage times are related by a detailed fluctuation theorem. This result descends from a symmetry of clockwise and counterclockwise trajectories, which combines time reversal with swapping portions of the trajectories. We then relate the entropy produced along the cycle with the entropy production of the whole system using large deviation theory. Our results reveal a novel symmetry in stochastic systems, of potential broad applicability in non-equilibrium physics.
\end{abstract}
\maketitle
\end{CJK*}

\maketitle

\section{Introduction}

First-passage times are of central importance in the theory of stochastic processes \cite{redner2001guide} and find widespread applications in physics and biology, including in reaction rate theory \cite{hanggi1990reaction}, spin models \cite{bray2013persistence}, epigenetics \cite{aurell2002epigenetics}, and neural dynamics \cite{burkitt2006review}. Our focus in this paper are first-passage times in physical systems, where stochasticity originates from interactions with a heat reservoir. Such systems present universal statistical properties, whose understanding is the subject of stochastic thermodynamics \cite{peliti2021stochastic}. Some properties of first-passage times have been considered in this context. For example, it has been shown that irreversibility sets a bound on the relative fluctuations of first-passage times \cite{gingrich2017fundamental,hiura2021kinetic,pal2021thermodynamic,raghu2025thermodynamic}. 

The statistical behavior of key observables in stochastic thermodynamics drastically depend on whether the system is at equilibrium or not. A main reason is that systems at thermodynamic equilibrium present an invariance under time reversal; this symmetry constrains the behavior of thermodynamic observables. However, this general idea cannot be naively applied to first-passage times. For example, the first-passage time $\tau_{xy}$ from a certain state $y$ to a certain state $x$ of a stochastic system is not generally related with $\tau_{yx}$, even if the system is at thermodynamic equilibrium. The reason is that $\tau_{xy}$ and $\tau_{yx}$ are statistical quantities pertaining to two different ensembles, in which either $x$ or $y$ are absorbing states, so that one of them is not trivially connected with the other by an operation of time reversal.

\begin{figure}
    \centering
    \includegraphics[width=0.8\textwidth]{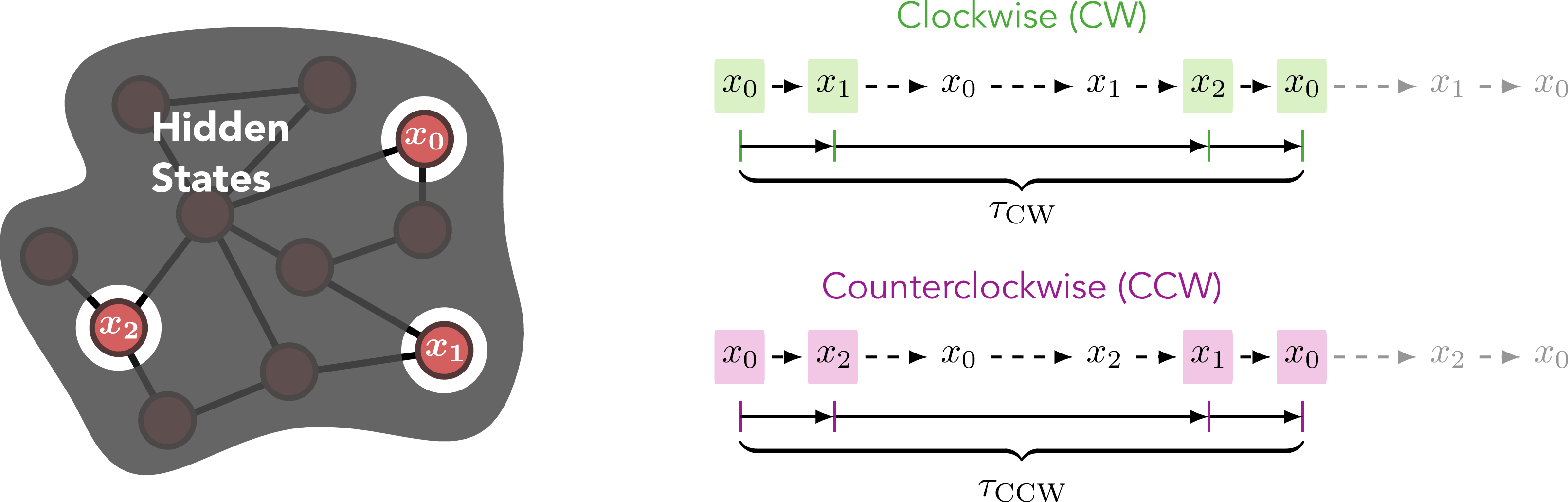}
    \caption{Clockwise (CW) and counterclockwise (CCW) trajectories over a set of three visible states, $x_0$, $x_1$, and $x_2$, among many hidden states in the system. In the right representation, the CW (top) and CCW (bottom) trajectories are highlighted. }
    \label{fig:intro}
\end{figure}

In this manuscript, we explore a symmetry that instead involves sums of first-passage times along a closed path in a stationary stochastic system. Specifically, we consider a sequence of $N>2$ distinct states $x_0, x_1 \dots x_{N-1}$, with $x_N \equiv x_0$ to close the path. We define the cyclic first-passage time as:
\begin{equation}
\tau_{\CW}=\sum_{j=1}^{N} \tau_{x_j,x_{j-1}} \, ,
\end{equation}
where the subscript ``CW'' stands for clockwise. Similarly, we define the counterclockwise cyclic first-passage time as $\tau_{\CCW}=\sum_{j=1}^{N} \tau_{x_{j-1},x_{j}}$ (see Fig.~\ref{fig:intro}). We shall also call these quantities ``clockwise time'' and ``counterclockwise time'' for shortness. From a practical point of view, measuring clockwise and counterclockwise times only requires observing the $N$ chosen states. This means that these quantities are measurable also in situations where only a small subset of a system's states is visible, while the remaining states are hidden.

As for individual first-passage times, CW and CCW cyclic first-passage are not related by a simple time-reversal operation, in general. However, we shall see that, at thermodynamic equilibrium, $\tau_{\rm CW}$ and $\tau_{\rm CCW}$ follow the same probability distribution, independently of the system and of choice of the states composing the path. Additionally, out of equilibrium, the distributions of $\tau_{\rm CW}$ and $\tau_{\rm CCW}$ are connected by a detailed fluctuation theorem. In stochastic thermodynamics, fluctuation theorems often descend from a time-reversal symmetry at the trajectory level \cite{chetrite2008fluctuation,harris2007fluctuation,seifert2012stochastic,peliti2021stochastic}. In our case, we shall see that the symmetry relating CW and CCW trajectories is related with time reversal, but requires cutting and pasting different portions of each trajectory. Besides its fundamental relevance, our finding implies that a difference between the distributions of $\tau_{\CW}$ and $\tau_{\CCW}$ can be taken as evidence that the system is out of thermodynamic equilibrium. We make this link explicit by deriving a direct connection between the statistics of cyclic first-passage times and those of the entropy production rate. 

The paper is organized as follows. Section~\ref{sec:dft} presents the detailed fluctuation theorem for cyclic first passage times. Section~\ref{sec:symmetry} explains the symmetry between clockwise and counterclockwise trajectories that underlies this detailed fluctuation theorem. Section~\ref{sec:ldt} connects the scaled cumulants of the entropy production rate with those of cyclic first passage times, using tools of large deviation theory. Section~\ref{sec:enzymatic} presents an application of our theory to a model of an enzymatic cycle. Section~\ref{sec:discussion} presents conclusions and perspectives.

\section{Detailed fluctuation theorem for cyclic first-passage times}\label{sec:dft}

We consider a sequence of $N>2$ states explored by a stationary stochastic system. The theory that follows is valid for both discrete and continuous phase space; we only require the dynamics to be Markovian. We introduce the joint distribution $p_{\CW}(\tau,s)$ of observing a clockwise time $\tau_{\CW}=\tau$ along the sequence while producing a total entropy $s$. We define $p_{\CCW}(\tau,s)$ in a similar way. Our first main result is a detailed fluctuation theorem:
\begin{equation}\label{eq:detailed}
\frac{p_{\CW}(\tau,s)}{p_{\CCW}(\tau,-s)} = e^{s} .
\end{equation}
Here and in the following, we choose units so that the Boltzmann constant and the temperature are both equal to one. Eq.~\eqref{eq:detailed} implies that, at equilibrium, the marginalized clockwise and counterclockwise distributions are identical: 
\begin{equation}
p_{\CW}(\tau)=p_{\CCW}(\tau) \qquad \mbox{(equilibrium)}  \; ,
\end{equation}
while in general the two distributions are related by 
\begin{equation}
    p_{\CW}(\tau) = \int ds ~p_{\CCW}(\tau,s) e^{-s} \; .
    \label{eq:pCWnoneq}
\end{equation}

To prove this result, we first introduce the notation $\bm{x}$ for a stochastic trajectory, and indicate its path probability with $\mathcal{P}(\bm{x})$. Given a sequence of $N>2$ states, we define a ``clockwise trajectory'' as a trajectory associated with a clockwise time, and similarly for counterclockwise trajectories. Any clockwise or counterclockwise trajectory starts and ends at $x=x_0$, by definition. We now express the joint probability $p_{\rm CW}(\tau, s)$ in terms of path probabilities:
\begin{equation}
    p_{\rm CW}(\tau, s) = \int_{\CWX} \mathcal{D}\bm{x}\, \mathcal{P}(\bm{x}) \delta_\tau \delta_{s}
\end{equation}
where $\CWX$ is the set of all clockwise trajectories,  $\delta_\tau$ is a short-hand notation for $\delta(\tau(\bm{x})-\tau)$, where $\tau(\bm{x})$ is the duration of the trajectory, and similarly $\delta_s=\delta(s(\bm{x})-s)$ where $s(\bm{x})$ is the entropy produced along the trajectory.

Our proof is centered on the existence of a combined operator $\hat{T}\hat{C}$ that maps a clockwise trajectory into a counterclockwise one. Here, the operator $\hat{T}$ represents time reversal, while $\hat{C}$ is an operator that preserves both path probability and entropy production, and that we will present in detail in the next Section.
Importantly, the operation $\hat{T}\hat{C}$ is an involution, meaning that $\hat{T}\hat{C}\hat{T}\hat{C}=\hat{I}$, where $\hat{I}$ is the identity operator on $\CWX$ (or $\CCWX$, the analogous of $\CWX$ for counterclockwise trajectories). By applying the $\hat{T}\hat{C}$ operator, we have:
\begin{eqnarray}
    p_{\rm CW}(\tau, s) &=& \int_{\CWX} \mathcal{D}\bm{x}\, \frac{\mathcal{P}(\bm{x})}{\mathcal{P}(\hat{T}\hat{C}\bm{x})} \mathcal{P}(\hat{T}\hat{C}\bm{x})  \delta_\tau \delta_{s} = \nonumber \\
    &=& \int_{\CCWX} \mathcal{D}\bm{x}\, \mathcal{P}(\bm{x}) e^{-s(\bm{x})}  \delta_\tau \delta_{-s} = p_{\rm CCW}(\tau,-s) e^{s} \;.
\end{eqnarray}
We obtained this result by the following steps. In the first line, we multiplied and divided by $\mathcal{P}(\hat{T}\hat{C}\bm{x})$. In the second line, we changed the integration domain into the set of counterclockwise trajectories. and used the irreversibility relation $\mathcal{P}(\hat{T}\hat{C}\bm{x})/\mathcal{P}(\bm{x}) = e^{-s(\bm{x})}$ \cite{peliti2021stochastic}.

\section{The symmetry leading to the detailed fluctuation theorem}\label{sec:symmetry}

We now properly introduce the operator $\hat{C}$ and its properties. Given a set of $N$ visible states, we represent trajectories as discrete sequence of visited states among the visible ones. These sequences always start and end in $x_0$, and have variable length $l$. We call $K$ the set of such cyclic sequences. A CW trajectory has therefore the form
\begin{equation*}
    \bm{x} = \big\{ \overbrace{x_0, \underbrace{\dots}_{\not\ni x_1} x_1}^{x_0\to x_1}, \underbrace{\dots}_{\not\ni x_2} x_2, \dots \overbrace{x_{N-1}, \underbrace{\dots}_{\not\ni x_0} x_0}^{x_{N-1}\to x_0} \big\} \in {\rm CW}_x \;,
\end{equation*}
A clockwise trajectory can visit $x_0$ several times. We also define the sets $\OCWX$ and $\OCCWX$. A trajectory belongs to $\OCWX$ ($\OCCWX$) if its time-reverse is a CW (CCW) trajectory, i.e., if it belongs to $\CWX$ ($\CCWX$). Our goal is to establish a relation between the sets $\CWX$ and $\OCCWX$. There exist trajectories that are both in $\CWX$ and in $\OCCWX$, e.g., for $N=3$,
\begin{equation}
    \bm{x} = \{ x_0, x_1, x_2, x_3, x_0\} \in \CWX, \OCCWX \nonumber \;,
\end{equation}
but this is not always the case, e.g.,
\begin{equation}
    \bm{x} = \{ x_0, x_1, x_2, x_3, x_0, x_2, x_0 \} \in \OCCWX
    \;, \; \bm{x}\notin \CWX \nonumber \;.
\end{equation}
This means that, in general, CCW trajectories cannot be constructed from CW trajectories by just reversing the time.

Given a generic set of trajectories $S$, we define as $S$-minimal trajectory a trajectory belonging to $S$ that does not contain any subsequence of length $l>1$ which also belongs to $S$. We define a $K$-minimal trajectory as a ``block'', i.e., a block is a trajectory that only visits $x_0$ at its start and end. We call $n$ the number of blocks in a trajectory, so that the number of intermediate $x_0$s in the given trajectory is equal to $n-1$. Furthermore, we define as ``minimal cycle'' a $\CWX$-minimal trajectory and indicate it as $\{x_0, B_x, x_0\}$. We remark that a minimal cycle may or may not be a block. A minimal cycle is a $\OCCWX$-minimal trajectory as well. In fact, the states in a minimal cycle are visited in the CW order forward in time and in the CCW order when moving backward in time. Furthermore, when moving backward in time, the minimal cycle terminates as soon as it completes the CCW cycle, thus meeting the conditions for being a CCW trajectory.

A CW trajectory can contain an arbitrary number of blocks, but it must contain a unique minimal cycle at its end, i.e., it must be of the form
\begin{equation}
    \bm{x} = \{ x_0, \dots, x_0, B_x, x_0 \} \in \CWX \; ,
\end{equation}
since it has to end after having completed the cycle. Similarly, a $\overline{\rm CCW}$ trajectory must contain a unique minimal cycle at its beginning:
\begin{equation}
      \bm{x} = \{ x_0, B_x, x_0, \dots, x_0 \} \in \OCCWX \; .
\end{equation}
CW trajectories that are themselves minimal cycles belong to both $\CWX$ and $\OCCWX$.

We now introduce cut operators $\hat{C}_i$, $i=1\dots n-1$ acting on a trajectory containing $n-1$ intermediate $x_0$s. The cut operator $\hat{C}_i$ swaps the two portions of the trajectory before and after the $i$-th intermediate $x_0$. As an example,
\begin{eqnarray}
    \{x\} &=& \{x_0, \dots, x_0, B_x, x_0\} \in \CWX \nonumber \\
    \hat{C}_1 \{x\} &=& \{x_0, B_x, x_0, \dots, x_0\} \in \OCCWX\, .
\end{eqnarray}
Each operator $\hat{C}_i$ preserves the path probability. To see that, we call ${\bm x}^A$ and $\bm{x}^B$ the two portions of trajectory before and after the time at which the $i$-th intermediate $x_0$ is reached. Since the dynamics is Markovian, the path probability factorizes as
\begin{equation}\label{eq:pathsplit}
\mathcal{P}(\bm{x}|x_0)=\mathcal{P}(\bm{x}^A|x_0)\mathcal{P}(\bm{x^B}|x_0)\;.
\end{equation}
The invariance of the path probability under application of $\hat{C}_i$ trivially follows from the commutative property of the product in the right hand side of Eq.~\eqref{eq:pathsplit}.

The family of cut operators $\hat{C}_i$, $i=1\dots n-1$ maps a trajectory into $n-1$ trajectories of the same length. This relation defines an equivalence class, in the sense that each of the $n$ trajectories (including the original one) is mapped into all the others by one of the $\hat{C}_i$s.

\begin{figure*}[htb]
    \centering
    \includegraphics[width=0.9\textwidth]{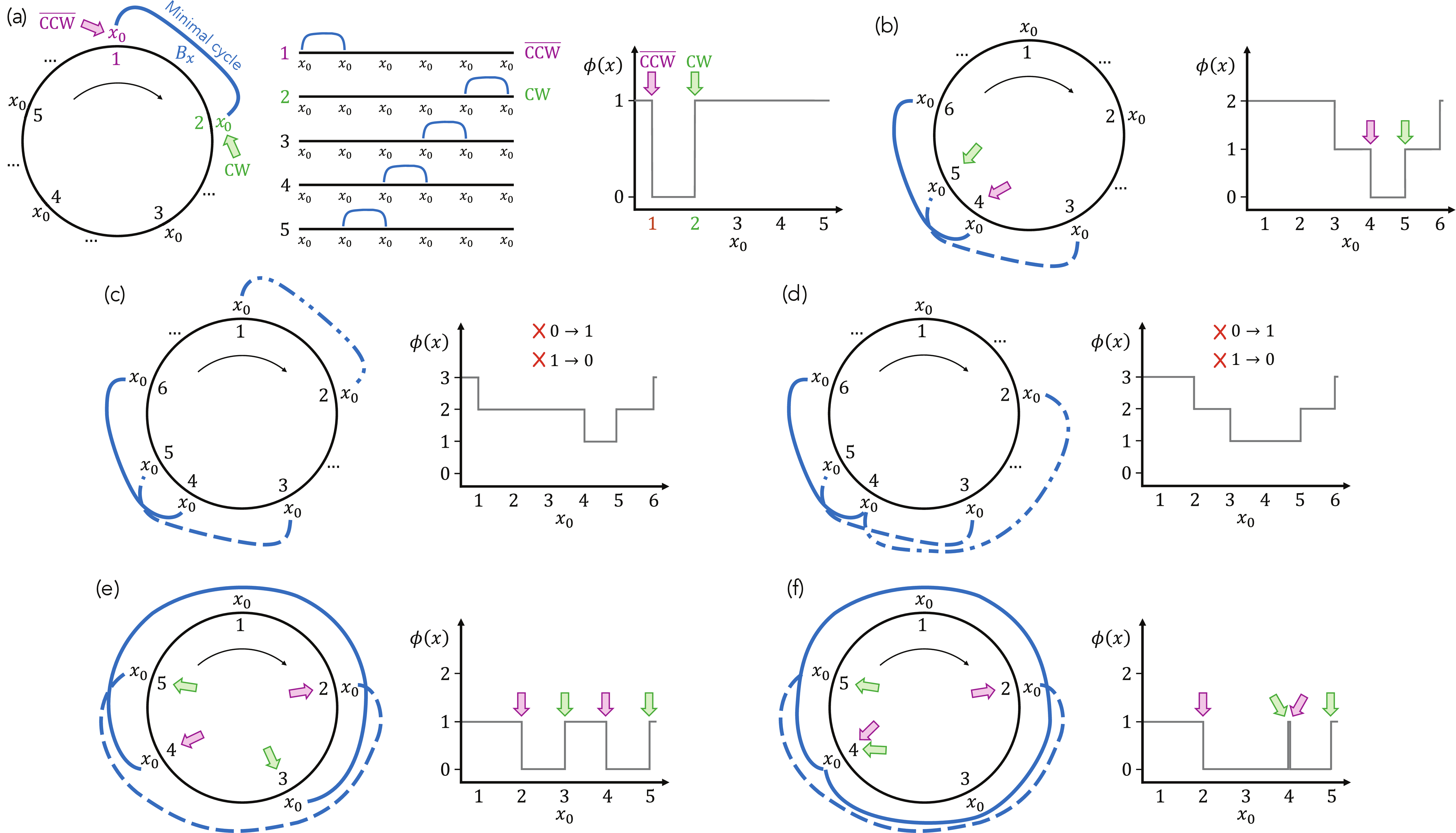}
    \caption{Equivalence class of cyclic trajectories in six examples of increasing complexity. (a) Example of a trajectory including a single minimal cycle. The two $x_0$s whose cuts lead to a ${\rm CW}$ and a $\overline{\rm CCW}$ trajectory are marked with arrows. The five trajectories associated with all possible cuts are shown in the right, alongside the position of the minimal cycle. The function $\phi(x)$ associated with the equivalence class is also shown. (b) Two overlapping minimal cycles are present and two different cuts lead to a CW and a $\overline{\rm CCW}$ trajectory. (c-d) Two cases in which none of the cuts lead to either ${\rm CW}$ or $\overline{\rm CCW}$ trajectories, due to the presence of multiple non-overlapping minimal cycles. (e-f) Examples in which multiple ${\rm CW}$ and $\overline{\rm CCW}$ trajectories are present in the same equivalence class. In example (f), one of the trajectories belongs to both ${\rm CW}$ and $\overline{\rm CCW}$. In all examples, the number of ${\rm CW}$ trajectories in each equivalence class is equal to the number of $\overline{\rm CCW}$ trajectories.}
    \label{fig:graphic}
\end{figure*}

We graphically represent this equivalence class by an oriented ring of states, see Fig.~\ref{fig:graphic}a. In this representation, each trajectory in the equivalence class corresponds to a choice of a particular $x_0$ as a starting and end point along the ring. We represent minimal cycles as arches on top of the ring.

We shall prove that each equivalence class contains the same number of ${\rm CW}$ and $\overline{\rm CCW}$ trajectories. Thanks to this property, one can define a bijection between $\CWX$ and $\OCCWX$ by arbitrarily pairing ${\rm CW}$ and $\overline{\rm CCW}$ trajectories in each of the equivalence classes. Calling $\hat{C}$ the choice of cut corresponding to this pairing, then $\hat{T}\hat{C}$ is a bijection between the sets $\CWX$ and $\CCWX$. 

To prove this property, we define a function $\phi(x)$ that depends on the state $x$ along the ring and counts the number of ``surviving'' minimal cycles if a trajectory is created by opening the ring at position $x$ (see Fig.~\ref{fig:graphic}a). By definition, $\phi(x)$ is a periodic function.  We notice that only one minimal cycle can start (or end) at each $x_0$. The reason is that one of the two minimal cycles starting/ending at the same point would contain the other, and therefore it would not be minimal. This property implies that $\phi(x)$ can only change by one unit at a step, i.e., from $n\to n\pm 1$. It is however possible that one minimal cycle ends at the same point in which another (or the same) minimal cycle starts. We represent these events as ``spikes'' in the function $\phi(x)$, i.e. transitions from $n\to n+1$ immediately followed by transitions from $n+1\to n$, see Fig~\ref{fig:graphic}f.

We now note that transitions of $\phi(x)$ from $1$ to $0$ mark the end of the only surviving minimal cycle in the trajectory. Therefore, such points $x=x_0$ are associated with a $\OCCWX$ trajectory. Similarly, transitions from $0$ to $1$ denote the starting points a unique minimal cycles and, as such, mark the $x_0$s associated with a $\rm CW$ trajectory. Since $\phi(x)$ is a periodic function and can only change by one unit at a step, the numbers of transitions $1\to 0$ and $0\to 1$ must be equal. This directly implies that each equivalence class contains equal numbers of $\rm CW$ and $\overline{\rm CCW}$ trajectories, as we were meant to show. Examples of equivalence classes and corresponding function $\phi(x)$ are shown in Figs.~\ref{fig:graphic}a-f. 

In Appendix~\ref{app:renewal}, we also provide an alternative proof of Eq.~\eqref{eq:detailed} based on renewal theory.

\section{Cyclic first-passage times and entropy production rate}\label{sec:ldt}

We now assume to know the probability $p_{\CW}(\tau,s)$, or equivalently $p_{\CCW}(\tau,s)$, for a given cycle. Our goal is to use this information to compute the large deviation function of the total entropy production.

We consider a long trajectory that completes the cycle $N_C$ times, and define the intensive variables
\begin{eqnarray}
j_{\tilde{s}}=\frac{1}{N_C}\sum_{i=1}^{N_C} s^{(i)} \qquad \textrm{and} \qquad j_\tau=\frac{1}{N_C}\sum_{i=1}^{N_C} \tau^{(i)} \; ,
\end{eqnarray}
where $s^{(i)}$ and $\tau^{(i)}$ are the entropy and time associated with the $i$-th cycle, respectively. Each sum is composed of i.i.d. random variables, although their terms are pairwise correlated. Therefore, the joint scaled cumulant generating function is given by
\begin{eqnarray}
\Phi(k_s,k_\tau) = \lim_{N_C\rightarrow\infty}\frac{1}{N_C}\ln \left\langle e^{N_C[k_s j_{\tilde{s}}+ k_\tau j_\tau]} \right\rangle = \ln \left\langle e^{k_s s + k_\tau \tau} \right\rangle \;.
\end{eqnarray}
Also, the variables $j_{\tilde{s}}$ and $j_\tau$ obey a large deviation principle, meaning that $P(j_{\tilde{s}},j_\tau)\asymp e^{-N_C I(j_{\tilde{s}},j_\tau)}$. The Gartner-Ellis theorem states that the rate function can be obtained by a Legendre transform \cite{ellis2007entropy}:
\begin{equation}
I(j_{\tilde{s}}, j_\tau) =\sup_{k_s,k_\tau\in \Re} [k_s j_{\tilde{s}} + k_\tau j_\tau -  \Phi(k_s,k_\tau) ] \;.
\label{eq:Isup}
\end{equation}
In this case, we fix $T = \sum_{i=1}^{N_C} \tau^{(i)}$ and introduce the new random variables $n_C = N_C/T$ and $j_s = j_{\tilde{s}} N_C/T$. By changing of variables in the probability distribution and using that the Jacobian does not contribute to the leading exponential order, 
we find that the joint large-deviation function $J(j_s, n_C)$ is given by
\begin{equation}
J(j_{s}, n_C) = n_C I(j_{s}/n_C, 1/n_C) 
\label{eq:Jdef}
\end{equation}
in terms of these new variables. We now use the contraction principle of large deviation theory to obtain
\begin{equation}
J(j_{s})=\inf_{n_C \in \Re^+} J(j_{s},n_C) .
\label{eq:Jsup}
\end{equation}
At this point, we can interpret $J(j_{s})$ as the large-deviation function associated with the empirical entropy production rate, since we marginalized over $n_C$.

By combining Eqs.~\eqref{eq:Isup}, \eqref{eq:Jdef}, and \eqref{eq:Jsup}, we have 
\begin{align}\label{eq:jjs}
J(j_{s}) = \inf_{n_C} n_C I(j_{s}/n_C,1/n_C) = \inf_{n_C} n_C \sup_{k_s,k_\tau} \left[\frac{k_s j_{s}+ k_\tau}{n_C} -  \Phi(k_s,k_\tau)\right] .
\end{align}
Assuming convexity, we swap the order of the suprema and rewrite Eq.~\eqref{eq:jjs} as 
\begin{equation}
J(j_{s})= \sup_{k_s} \left[k_s j_{s} - \inf_{n_C, k_\tau}\left(  n_C\Phi(k_s,k_\tau)- k_\tau  \right)\right] .
\end{equation}
Again from the Gartner-Ellis theorem, the scaled cumulant generating function of the empirical entropy production rate is expressed by
\begin{equation}\label{eq:gener}
\psi(k_s) = \lim_{T \rightarrow \infty}\frac{1}{T} \ln \left\langle e^{T k_s j_s}\right\rangle = \inf_{n_C,k_\tau} \left( n_C \Phi(k_s,k_\tau) - k_\tau \right) \; .
\end{equation}
The extremality conditions are
\begin{eqnarray}
\Phi(k_s,k_\tau^{\inf})=0 \qquad \mbox{or}\qquad n^{\inf}=0 \;, \qquad \textrm{and} \qquad
n^{\inf}\partial_{k_\tau} \Phi(k_s,k_\tau)|_{k_\tau=k_\tau^{\inf}}=1 \; .
\label{eq:conditions}
\end{eqnarray}
We focus on the first of these two conditions. Substituting this back into Eq.~\eqref{eq:gener}, we obtain the remarkably simple relationship
\begin{equation}\label{eq:geners}
\psi(k_s)=-k_\tau^{\inf} .
\end{equation}
We further assume that the non-trivial condition $n^{\inf}>0$ always holds (otherwise the second condition in Eq.~\eqref{eq:conditions} cannot be satisfied). This implies that $k_\tau^{\inf}$ can be obtained from the following relationship
\begin{equation}\label{eq:res_prel}
\left\langle e^{k_s s + k_\tau^{\inf} \tau} \right\rangle = 1 \;.
\end{equation}
Or, equivalently,
\begin{equation}\label{eq:res1}
\left\langle e^{k_s s - \psi(k_s) \tau} \right\rangle =1 .
\end{equation}
Equations~\eqref{eq:res_prel} and \eqref{eq:res1} are our second main result. The scaled cumulants of the entropy production rate can be obtained by expanding Eq.~\eqref{eq:res1} around $k_s=0$, and imposing that each term vanishes (since the function is equal to a constant). At the first order we obtain
\begin{equation}
\langle j_s \rangle = \mu_s = \frac{\langle s \rangle}{\langle \tau \rangle} \;.
\label{eq:meanLD}
\end{equation}
The second order term yields the scaled variance
\begin{eqnarray}
\sigma_s = \left\langle \left(j_s - \mu_s \right)^2 \right\rangle = \frac{d^2}{d k_s^2} \psi(k_s)|_{k_s=0} =\frac{1}{\left \langle \tau \right\rangle}
\left\langle\left(s - \frac{\langle s\rangle}{\langle \tau\rangle} \tau\right)^2 \right\rangle = \frac{1}{\left \langle \tau\right\rangle}
\left\langle\left(s- \mu_s \tau\right)^2 \right\rangle .
\label{eq:varLD}
\end{eqnarray}
Finally, we compute the third central moment:
\begin{gather}
    \langle (j_s - \mu_s)^3 \rangle = \frac{1}{\langle \tau \rangle} \bigg[ \left\langle \left( s - \mu_s \tau \right)^3 \right\rangle
    - 3 \sigma_s \left\langle \tau \left( s - \mu_s \tau \right) \right\rangle \bigg] \;.
    \label{eq:thirdLD}
\end{gather}
We remark that we did not specify whether these $N_C$ cycles are performed clockwise or counterclockwise. As a consequence, we can equivalently compute all averages in Eqs.~\eqref{eq:meanLD}, \eqref{eq:varLD}, and \eqref{eq:thirdLD} over clockwise or counterclockwise distributions. 

\section{Example: enzymatic cycle}\label{sec:enzymatic}

\begin{figure}[b]
    \centering
    \includegraphics[width=1\columnwidth]{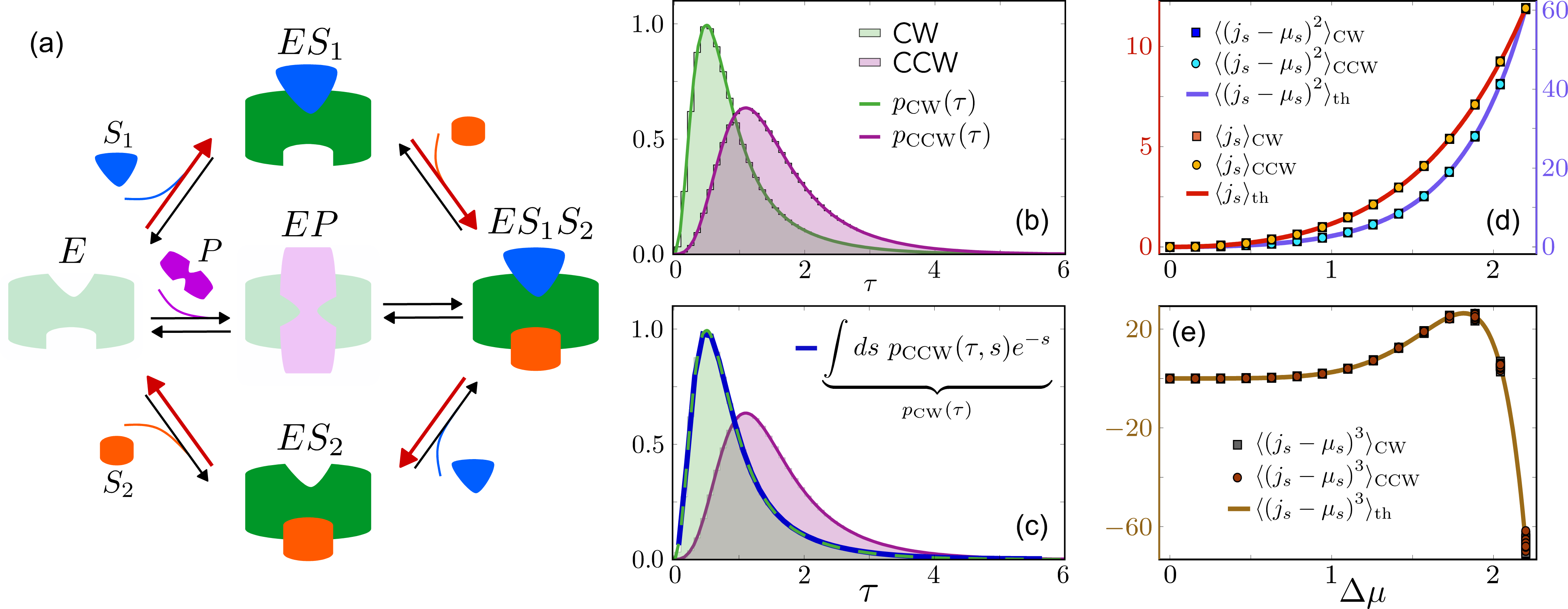}
\caption{
(a) Enzymatic cycle in which two substrates, $S_1$ and $S_2$, are converted into a product, $P$. Substrates and product concentrations are chemostatted. Only substrate-bound states are visible (not shaded) and the system is externally driven to cycle clockwise (red arrows). (b) Distribution of the first passage time associated with clockwise ($ES_1-ES_1S_2-ES_2-ES_1$) and counterclockwise ($ES_1-ES_2-ES_1S_2-ES_1$) visible cycles. (c) Numerical validation of the fluctuation theorem in Eq.~\eqref{eq:detailed}. The clockwise distribution is recovered from the counterclockwise statistics upon integration over all possible values of the entropy production. (d) Mean and variance of the entropy production rate estimated from the large-deviation argument in Eq.~\eqref{eq:res1} (applied to clockwise and counterclockwise cycles) coincide with their corresponding theoretical values. Here we denote by $\langle \cdot \rangle_{\rm CW}$ an average performed over the distribution of clockwise cycles (and similarly for $\langle \cdot \rangle_{\rm CCW}$), while $\langle \cdot \rangle_{\rm th}$ indicates the corresponding theoretical expression. (e) The same comparison is shown for the third central moment. Numerical values have been obtained from $20$ different realizations for each cycle direction (clockwise and counterclockwise), all reported in the figure. Energies ($\epsilon_i$) and energetic barriers ($B_{ij}$) are: $\epsilon_E = 0, \epsilon_{ES_1} = 0.023, \epsilon_{ES_1S_2} = 0.20, \epsilon_{ES_2} = 0.023, \epsilon_{EP} = 0.19, B_{E,ES_1} = 0.24, B_{E,ES_2} = 0.28, B_{E,EP} = 0.27, B_{ES_1,ES_1S_2} = 0.28, B_{ES_2,ES_1S_2} = 0.23, B_{ES_1S_2,EP} = 0.26$. We take $k_B T = 1$, $[S_1] = [S_2] = [P] = 1$, and the nonequilibrium driving equal to $\Delta\mu = 2$ in panels (a-c).
}
\label{fig:FT_val}
\end{figure}

We test our results using a model of an enzymatic cycle, see Fig.~\ref{fig:FT_val}a.. We consider an enzyme $E$ that can bind two different substrates, $S_1$ and $S_2$, thereby forming two distinct complexes, $ES_1$ and $ES_2$, with rates $k_{E\to ES_1}$ and $k_{E\to ES_2}$, respectively. These complexes can further bind to the other substrate, thus forming the trimer $ES_1S_2$. The enzyme catalyzes the conversion of both bound substrates into a product $P$, i.e., $ES_1S_2$ transforms into $EP$ with a rate $k_{ES_1S_2\to EP}$. This product is then released into the environment, thereby freeing the enzyme. For thermodynamic consistency, we take all  reactions to be reversible, i.e., if a chemical rate connecting any two states $i$ and $j$, $k_{i\to j}$, is different from zero, then $k_{j\to i} \neq 0$. The entire set of chemical reactions is 
\begin{align}
    E + S_1 &\leftrightharpoons ES_1 &\quad E + S_2 &\leftrightharpoons ES_2 \nonumber \\
    ES_1 + S_2 &\leftrightharpoons ES_1S_2   &\quad ES_2 + S_1 &\leftrightharpoons ES_1S_2 \nonumber \\
    E + P &\leftrightharpoons EP &\quad ES_1S_2 &\leftrightharpoons EP \nonumber
\end{align}
with the transition rates parametrized as:
\begin{equation}
    k_{i\to j} = e^{-(B_{ij} - \epsilon_i)} \;.
\end{equation}
Here, $\epsilon_i$ is the energy of the state $i$, and $B_{ij} = B_{ji}$ the energetic barrier associated with the interconversion of $i$ into $j$. We consider the concentrations of substrates and product to be chemostatted at their equilibrium values, i.e., $[S_1] [S_2]/[P] = 1$ in this case, for simplicity. This condition implies that, in the absence of any additional driving, the system satisfies detailed balance, and $p_{\rm CW}(\tau) = p_{\rm CCW}(\tau)$. We further consider the presence of a nonequilibrium driving $\Delta\mu$ that induces the enzyme to preferentially bind first $S_1$ and then $S_2$, and unbind first $S_2$ and then $S_1$. In Fig.~\ref{fig:FT_val}a, the red arrows indicate the rates modified by the driving as
\begin{equation}
    k_{i\to j} \to k^{\rm neq}_{i\to j} = k_{i\to j} e^{\Delta\mu} \;.
\end{equation}
We consider the scenario in which the enzyme can be observed only when bound to (one or both) substrates, see Fig.~\ref{fig:FT_val}a. We track clockwise and counterclockwise first-passage cycles in this reduced 3-state system and record their associated completion times at stationarity. We thereby compute and compare $p_{\rm CW}(\tau)$ and $p_{\rm CCW}(\tau)$, see Fig.~\ref{fig:FT_val}b. As expected, clockwise cycles are completed, on average, in shorter times out of equilibrium. 

Since we are dealing with a discrete-state system, given a cyclic trajectory visiting $N$ states, at stationarity, we compute the entropy by the expression \cite{peliti2021stochastic}:
\begin{equation}
    s(\bm{x}) = \sum_{j=1}^N \log \frac{k_{x_{j-1} \to x_j}}{k_{x_j \to x_{j-1}}} \;.
\end{equation}

We find excellent agreement between simulations and the fluctuation theorem embodied in
Eq.~\eqref{eq:pCWnoneq}, see Fig.~\ref{fig:FT_val}c. We now test our ability to infer the statistics of the entropy production rate $j_s$ from the entropy produced during clockwise or counterclockwise cycles. The expressions of mean, variance, and third central moment are respectively shown in Eqs.~\eqref{eq:meanLD}, \eqref{eq:varLD}, and \eqref{eq:thirdLD}. In Fig.~\ref{fig:FT_val}d-e, we compare the numerical results obtained by these formulas (both considering clockwise and counterclockwise cycles) with their exact analytical values, computed with the method detailed in \cite{busiello2022hyperaccurate,padmanabha2023fluctuations}. We find once again an excellent agreement that validates our theoretical results.

\section{Discussion}\label{sec:discussion}

In this work, we uncovered a nonequilibrium symmetry of cyclic first-passage times among an arbitrary subset of states of a stochastic system. This symmetry leads to a detailed fluctuation theorem that relates the timing of such cycles with the entropy dissipated in the process. Our approach makes minimal assumptions on the underlying dynamics and requires observing an arbitrary subset of $N>2$ states. These features make our approach particularly apt to probe the nonequilibrium behavior of biological stochastic systems that operate through cyclic processes.

Experimental advancements have made it possible to measure first-passage times at the molecular level \cite{grebenkov2018towards,berezhkovskii2021distributions,broadwater2021first}. In this context, our result can be applied to identify the main dissipative macroscopic cycles in a system, by comparing clockwise and counterclockwise trajectories between different subsets of visible states. Additionally, the connection we revealed between statistics of cyclic first passage times and entropy production rate complements previous efforts to infer entropy production from partial information \cite{li2019quantifying,skinner2021improved,harunari2022What,van2022thermodynamic}. However,  estimating the dissipation associated with each cycle might be experimentally challenging, and potentially require additional information, such as the amount of consumed ATP in chemically active systems. Nevertheless, our alternative approach exploits this additional information to infer, in principle, the entire statistics of the total entropy production.

Our theory can be extended in several directions to facilitate experimental application. In particular, experimentally accessible states are often coarse-grained, including unresolved internal degrees of freedom. Extending the nonequilibrium symmetry uncovered here to such situations is a nontrivial problem that we leave for future investigations. Another promising extension is to consider intertwined clockwise and counterclockwise cycles, with the goal of using this nested structure to probe additional nonequilibrium features of a system.

\begin{acknowledgements}
DMB conducted part of this research while visiting the Okinawa Institute of Science and Technology (OIST) through the Theoretical Sciences Visiting Program (TSVP). All authors thanks the TSVP since part of initial discussions on this study were conducted during a TSVP workshop. DMB is funded by the STARS@UNIPD program through the project ``ActiveInfo''.
\end{acknowledgements}

\appendix

\section{Alternative derivation of the detailed fluctuation theorem}\label{app:renewal}

Here we use an alternative approach based on Laplace transform and renewal theory to prove the detailed fluctuation theorem for the cyclic first-passage-time distribution in Eq.~\ref{eq:detailed}. The cyclic first-passage probability is defined as the probability of sequential first-passage events, following the order $[x_0,x_1,\cdots,x_{N-1},x_0]$. Any two adjacent states in the sequence can be an observable subset of the whole set of states in the system and, as such, they are connected the first passage events as well. The probability of completing the cycle in a time $\tau$, conditioned on producing an entropy $s$, can be written as the (bivariate) convolution of first-passage distributions between adjacent states:
\begin{equation}
  p_{\mathrm{CW}}(\tau,s)=\bigstar_{j=1}^{N} f_{x_j,x_{j-1}}(\tau,s) \
  ,
\end{equation}
where $\star$ denotes convolution in $(\tau,s)$:
\[
  (f\star g)(\tau,s)=\int_0^{\tau}\!\!dt'\int_{-\infty}^{\infty}\!\!ds'\; f(t',s')\,g(\tau-t',\,s-s') \;.
\]
Taking the Laplace transform in $(\tau,s)$ (unilateral in $\tau$, bilateral in $s$) gives
\begin{equation}
  P_{\mathrm{CW}}(u,v)=\mathcal{L}_{\tau,s}\{p_{\mathrm{CW}}(\tau,s)\}
  =\prod_{j=1}^{N} F_{x_j,x_{j-1}}(u,v) \;.
\end{equation}
By renewal theory, first-passage and propagator Laplace transforms are related by
\begin{equation}
  F_{x_j,x_{j-1}}(u,v)=\frac{P_{x_j,x_{j-1}}(u,v)}{P_{x_j,x_j}(u,v)} \;,
\end{equation}
so that we have
\begin{equation}\label{eq:PCW-product}
  P_{\mathrm{CW}}(u,v)=\prod_{j=1}^{N}\frac{P_{x_j,x_{j-1}}(u,v)}{P_{x_j,x_j}(u,v)} \;.
\end{equation}
For the counterclockwise (CCW) cycle $[x_0,x_{N-1},\dots,x_1,x_0]$, this means
\begin{equation}\label{eq:PCCW-product}
  P_{\mathrm{CCW}}(u,v)=\prod_{j=1}^{N}\frac{P_{x_{j-1},x_j}(u,v)}{P_{x_{j-1},x_{j-1}}(u,v)} \;.
\end{equation}
We now use the fluctuation theorem for propagators:
\begin{equation}
  \frac{p_{x,y}(\tau,s)}{p_{y,x}(\tau,-s)}=e^{s}
  \qquad\Longleftrightarrow\qquad
  P_{x,y}(u,v)=P_{y,x}(u,1-v).
\end{equation}
Setting $x=y$ also yields $P_{x,x}(u,v)=P_{x,x}(u,1-v)$. Applying $P_{x_j,x_{j-1}}(u,v)=P_{x_{j-1},x_j}(u,1-v)$ and $P_{x_j,x_j}(u,v)=P_{x_j,x_j}(u,1-v)$ to \eqref{eq:PCW-product}, we have:
\begin{equation}
  P_{\mathrm{CW}}(u,v)=\prod_{j=1}^{N}\frac{P_{x_{j-1},x_j}(u,1-v)}{P_{x_{j-1},x_{j-1}}(u,1-v)}
  \;=\; P_{\mathrm{CCW}}(u,1-v).
\end{equation}
This is the detailed fluctuation theorem in the Laplace domain. Taking the inverse Laplace transforms for $u$ and $v$ directly yields Eq.~\eqref{eq:detailed}.

\end{document}